\documentclass[prl,twocolumn,groupedaddress]{revtex4-1}
\catcode`\@=11
\def\lsim{\mathrel{\mathpalette\@versim<}}
\def\gsim{\mathrel{\mathpalette\@versim>}}
\def\@versim#1#2{\vcenter{\offinterlineskip
\ialign{$\m@th#1\hfil##\hfil$\crcr#2\crcr\sim\crcr } }}
\catcode`\@=12

\catcode`@=12
\usepackage[export]{adjustbox}	
\usepackage{float}
\usepackage{caption}
\usepackage{subcaption}
\usepackage{epsfig,bbm}
\usepackage{amsmath}
\usepackage{slashed}
\usepackage{tikz}
\usepackage{bm}
\usepackage{amssymb}
\usepackage{mathtools}
\usepackage{graphicx}
\newcommand{\be}{\begin{equation}}
\newcommand{\ee}{\end{equation}}
\newcommand{\bea}{\begin{eqnarray}}
\newcommand{\eea}{\end{eqnarray}}

\begin{document}
\thispagestyle{empty}
\begin{flushright}
\end{flushright}

\begin{center}
\title{ Light inflaton completing Higgs inflation
}
\author{Hyun Min Lee}
\email[]{hminlee@cau.ac.kr}
\affiliation{Department of Physics, Chung-Ang University, Seoul
06974, Korea}
\begin{abstract}
We consider the extension of the Standard Model (SM) with a light inflaton where both unitarity problem in Higgs inflation and vacuum instability problem are resolved. The linear non-minimal coupling of the inflaton to gravity leads to a significant kinetic mixing between the inflaton and the graviton such that perturbative unitarity is restored up to Planck scale. 
We show the correlation between unitarity scale and inflationary observables in this model and discuss how the effective Higgs inflation appears. 
  
\end{abstract}
\maketitle
\end{center}

\section{Introduction}

Cosmic inflation \cite{beginning,beginning2} solves horizon problem and explains isotropy, homogeneity, flatness of the Universe, etc, in Big Bang cosmology (See, for instance, Ref.~\cite{burgess,models} for a recent review).
Observation of Cosmic Microwave Background (CMB) anisotropies \cite{planck0,planck,cmb} is well consistent with the existence of the early period of slow-roll inflation and quantum fluctuations during inflation seeds the large scale structure of the Universe. 


Predictions of large field  models for inflation can be sensitive to unknown high-scale physics unless there is a reason to have the power expansion of inflaton potential at any new physics scale under control such that the semi-classical approximation is justified during inflation. Identification of new physics scales depends on the inflaton field values and the kinetic terms during inflation, so  the power counting for the effective field theory should be applied with care to capture physics during inflation.

It is a legitimate question whether there is a way to obtain a viable inflation model with the known or testable particles and interactions at low energy. 
Higgs inflation \cite{Higgsinf} uses the SM Higgs boson as the inflaton but the validity of the inflaton potential from electroweak all the way to large field values during inflation is challenging. The non-minimal coupling of the Higgs boson to gravity must be of order $\xi\sim 10^4$, violating perturbative unitarity at $M_P/\xi$ in the SM vacuum, comparable to the Hubble scale $H\sim M_P/\xi$ during inflation \cite{unitarity}. But, new physics scale becomes field-dependent during inflation \cite{fedor} and gets larger to the Planck scale in the inflaton sector so a scale-invariant completion does not change the inflaton potential in the region in which the inflationary observables are evaluated \cite{fedor}. On the other hand, the new physics scale is saturated to $M_P/\sqrt{\xi}$ in the gauge sector, so the semi-classical expansion in powers of $\sqrt{\xi} H/M_P$ or the inflation potential might depend on unknown new physics entering at $M_P/\sqrt{\xi}\sim \sqrt{\xi} H$ \cite{burgess}. 

Furthermore, the quartic self-coupling of the Higgs boson runs to smaller but positive values at high energies, although its precise value depends on the top quark pole mass and the strong coupling at low energy \cite{vsb,add,sthreshold}. In this context, there is an interesting possibility that an inflection point in the Higgs potential at high energies can be used for inflation \cite{inflection}. However, if a large top quark pole mass is taken, the vacuum instability scale is much lower than the typical inflation scale in Higgs inflation. For instance, for $m_t=173.2\,{\rm GeV}$, together with $m_h=125\,{\rm GeV}$ and $\alpha_s(M_Z)=0.1183$, the vacuum instability scale is given by $\Lambda_I=4\times 10^{10}\,{\rm GeV}$ \cite{sthreshold}, so the Higgs boson would not be appropriate for a dominant component of the inflaton in this case.

In this paper, we propose a simple extension of the SM with a light singlet scalar field as a dominant component of the inflaton. We introduce non-minimal couplings of the singlet scalar to gravity at both linear ($\xi_1$) and quadratic ($\xi_2$) orders and discuss the roles of the singlet field as an Ultra-Violet (UV) completion of Higgs inflation \cite{gian,gian2,espinosa} as well as for solving the vacuum instability problem. A large linear non-minimal coupling with $\xi_1\sim \sqrt{\xi_2}$ allows for a significant kinetic mixing of the singlet field with the graviton, ensuring perturbative unitarity up to Planck scale. In this model, we show that the linear non-minimal coupling makes a crucial difference from Higgs inflation.

\section{Light singlet inflaton}

The most general Lagrangian with the SM Higgs boson $\phi$ and a light singlet sigma field $\sigma$ coupled to gravity  is the following,
\bea
\frac{{\cal L}}{\sqrt{-g}}= \frac{1}{2}M^2_P\, \Omega(\sigma,\phi) R - \frac{1}{2}(\partial_\mu\sigma)^2 -\frac{1}{2}(\partial_\mu\phi)^2 -V_J(\sigma,\phi) \nonumber \\  \label{fullag}
 \eea 
where the frame function and the scalar potential are given by
\bea
\Omega(\sigma,\phi) & =& 1+\frac{\xi_1\sigma}{M_P}+\frac{1}{M^2_P}(\xi_2\sigma^2+\xi_\phi\phi^2),  \label{frame}  \\ 
V_J(\sigma,\phi)&=&\frac{1}{2} m^2_\phi \phi^2+\frac{1}{4}\lambda_\phi \phi^4+ \frac{1}{2}m^2_\sigma 
\sigma^2 -\mu\sigma \phi^2 \nonumber \\
&& +\frac{1}{3}\alpha\sigma^3 +\frac{1}{2}\lambda_{\sigma\phi} \sigma^2\phi^2+\frac{1}{4}\lambda_\sigma\sigma^4. \label{fullpot}
\eea
This is a generalized form of the UV complete Higgs inflation proposed in Ref.~\cite{gian,espinosa}
or Starobinsky models for inflation in Ref.~\cite{general,gian2}.
We take the mass parameters for $\sigma$ in the above scalar potential (\ref{fullpot}), i.e. $m_\sigma, \mu, \alpha$,  to be of order the weak scale for a light inflaton. 
A similar model with a singlet scalar field to ours was considered in Ref.~\cite{oleg2} but the effect of the linear non-minimal coupling was not discussed. The detailed discussion on the light inflaton with a small non-minimal coupling, $\xi_2<1$ and $\xi_1=0$, can be found in Ref.~\cite{lightinflaton}.

Assuming $\xi_2, \xi_\phi>0$,
 we only have to impose the condition for stable gravity as
 \be
 \xi^2_1<4 (\xi_2+\xi_\phi \tau^2),  \label{stablegravity}
 \ee
with $\tau=\phi/\sigma$,
because otherwise the effective Planck mass squared could become negative during the cosmological evolution. Thus, eq.~(\ref{stablegravity}) leads to the upper bound on the linear non-minimal coupling $\xi_1$ for stable gravity in the entire field space.

From eq.~(\ref{fullag}), setting $M_P=1$ and performing the metric rescaling by $g_{\mu\nu}=g^E_{\mu\nu}/\Omega$  with $\Omega=1+\xi_1 \sigma+\xi_2\sigma^2+\xi_\phi\phi^2$,  
we get the Einstein frame Lagrangian as 
\bea
\frac{{\cal L}}{\sqrt{-g_E}}&=& \frac{1}{2} R(g_E) -\frac{1}{2\Omega}(\partial_\mu{\sigma})^2-\frac{3}{4}(\partial_\mu\log \Omega)^2 \nonumber \\
&&-\frac{1}{2\Omega}(\partial_\mu\phi)^2 - V_E(\sigma,\phi) \label{sigma-ein} 
\eea
where the Einstein frame potential is given by
$V_E(\sigma,\phi)=V_J/\Omega^2$.

\section{Unitarity and Higgs inflation}

 \begin{figure}
  \begin{center}
    \includegraphics[height=0.40\textwidth]{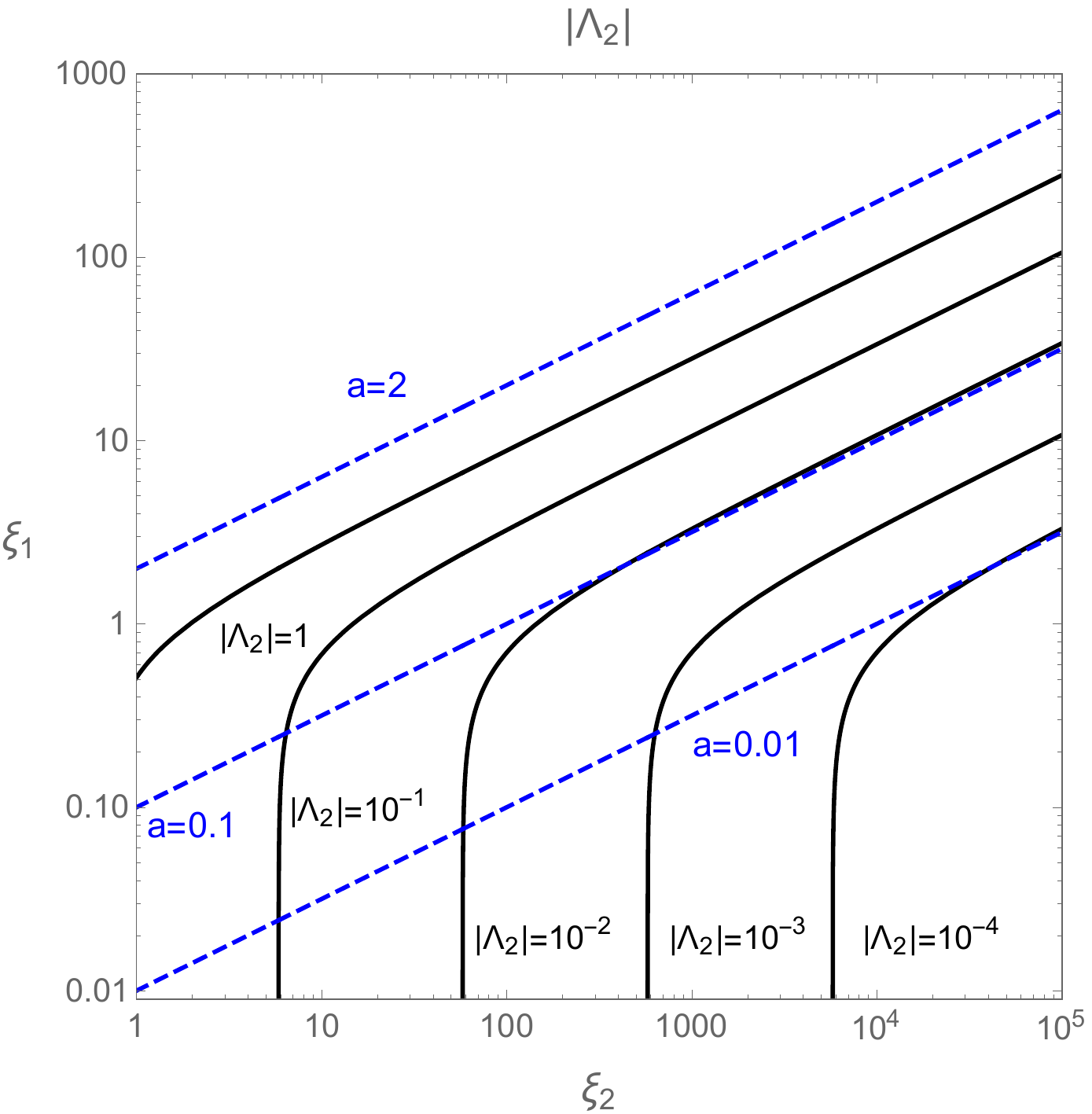}
  \end{center}
  \caption{Contours of $|\Lambda_2|$  in units of $M_P$ in the plane of $\xi_2$ and $\xi_1$ in black solid lines.  We overlaid in blue dashed lines the contours of $a=\xi_1/\sqrt{\xi_2}$.}
  \label{cutoff}
\end{figure}

We consider the unitarity scale in our model with the light inflaton and discuss the effective Higgs inflation after the sigma field is integrated out.
Near the true minimum with $\sigma,\phi\ll 1$, we get the quadratic kinetic terms in eq.~(\ref{sigma-ein}) approximated to
\bea
{\cal L}_{{\rm kin},0}= -\frac{1}{2} \Big(1+\frac{3}{2} \xi^2_1 \Big)(\partial_\mu\sigma)^2-\frac{1}{2}(\partial_\mu\phi)^2.
\eea
Then, redefining the sigma field by
\bea
{\tilde\sigma}= \Big(1+\frac{3}{2} \xi^2_1\Big)^{1/2} \sigma. \label{rescale}
\eea
we obtain the leading derivative interaction terms are
\bea
{\cal L}_{\rm int}&=& \frac{1}{\Lambda_1}\, {\tilde\sigma}(\partial_\mu{\tilde\sigma})^2+ \frac{1}{\Lambda^2_2}\, {\tilde\sigma}^2(\partial_\mu{\tilde\sigma})^2 +\frac{1}{\Lambda^2_3}\,\phi^2(\partial_\mu{\tilde\sigma})^2 \nonumber \\
&&+\frac{1}{\Lambda_4}\,{\tilde\sigma}(\partial_\mu\phi)^2 +\frac{1}{\Lambda^2_5}\, {\tilde\sigma}^2(\partial_\mu\phi)^2 +\frac{1}{\Lambda^2_6}\, \phi^2(\partial_\mu\phi)^2 \nonumber \\
&&-\frac{1}{\Lambda_7}\, \phi(\partial_\mu{\tilde\sigma})(\partial^\mu\phi)-\frac{1}{\Lambda^2_8}\, {\tilde\sigma}\phi (\partial_\mu{\tilde\sigma})(\partial^\mu\phi)+\cdots
\eea
where the ellipses are higher dimensional terms and the cutoff scales in the leading terms read
\bea
\Lambda_1&\equiv &\frac{2\Big(1+\frac{3}{2}\xi^2_1\Big)^{3/2}}{\xi_1(1+3\xi^2_1-6\xi_2)} , \\
|\Lambda_2|&\equiv &\frac{\sqrt{2}\Big(1+\frac{3}{2}\xi^2_1\Big)}{\Big|\xi^2_1\Big(1+\frac{3}{2}\xi^2_1\Big)+\xi_2 (1+3\xi^2_1-6\xi_2)\Big|^{1/2}}, \\
\Lambda_3&\equiv&\sqrt{\frac{2\Big(1+\frac{3}{2}\xi^2_1\Big)}{\xi_\phi(1+3\xi^2_1)}}, \\
\Lambda_4 &=&\frac{2\sqrt{1+\frac{3}{2}\xi^2_1}}{\xi_1} , \\
\Lambda_5&=&\sqrt{ \frac{2\Big(1+\frac{3}{2}\xi^2_1\Big)}{\xi^2_1+\xi_2}}, \\
|\Lambda_6|&=&\sqrt{\frac{2}{|\xi_\phi(1-6\xi_\phi)|}}, \\
\Lambda_7 &=&\frac{\sqrt{1+\frac{3}{2}\xi^2_1}}{3\xi_\phi\xi_1}, \\
\Lambda_8&=&\sqrt{\frac{1+\frac{3}{2}\xi^2_1}{6\xi_\phi\xi_2}}.
\eea
Therefore, taking $\xi_\phi={\cal O}(1)$ to avoid unitarity problem due to Higgs interactions such as $\Lambda_{6,7}$,  we can read off the minimum cutoff scales from $\Lambda_{1,2,5,8}$. In Fig.~\ref{cutoff}, we depict the minimum cutoff scale $|\Lambda_2|$  in units of $M_P$ in the plane of $\xi_2$ and $\xi_1$. Then, we find that the cutoff scale of order Planck scale requires a sizable linear non-minimal coupling, namely, $a={\cal O}(1)$ or $\xi_1\sim \sqrt{\xi_2}$.

Consequently, mass parameters in the scalar potential (\ref{fullpot}) are not constrained by unitarity, as far as they are below the scale of unitarity violation in Higgs inflation. Thus, the sigma field mass can be of order weak scale or lower, being consistent with the UV complete inflation, unlike the cases only with either $\xi_2$ \cite{gian} or $\xi_1$ \cite{espinosa}, where the new singlet scalar must be much heavier than weak scale to get a large VEV from the renormalizable scalar potential \cite{gian} or match the COBE normalization with the singlet mass term \cite{espinosa}.

In order to discuss the effective Higgs inflation,
we plug into eq.~(\ref{fullag}) the equation of motion for $\sigma$ with
 \be
 \sigma=\sqrt{\frac{-m^2_\sigma-\lambda_{\sigma\phi}\phi^2}{\lambda_\sigma}} \label{sigmaeq}
 \ee
 where the dimensionful interactions are ignored in the scalar potential (\ref{fullpot}). 
 As a result, we obtain the effective Lagrangian for the Higgs inflation from as
 \bea
 \frac{{\cal L}_{\rm eff}}{\sqrt{-g}}= \frac{1}{2} M^2_P\,\Omega_{\rm eff}(\phi)R -\frac{1}{2}(\partial_\mu\phi)^2 -V_{\rm eff}(\phi) 
 \label{Higgseff}
 \eea
 where the effective frame function and Higgs potential become
 \bea
 \Omega_{\rm eff}&=&1-\frac{\xi_2 m^2_\sigma}{\lambda_\sigma M^2_P}+\frac{\xi_1}{M_P}\sqrt{\frac{-m^2_\sigma-\lambda_{\sigma\phi}\phi^2}{\lambda_\sigma}}+\frac{\xi_{\phi,{\rm eff}} \phi^2}{M^2_P},   \label{efframe} \\
 V_{\rm eff}&=& \frac{1}{2} m^2_\phi \phi^2 + \frac{1}{4}\lambda_{\rm eff} \phi^4
 \eea
 with
 \bea
 \xi_{\phi,{\rm eff}} &\equiv&\xi_\phi-\frac{\lambda_{\sigma\phi}\xi_2}{\lambda_\sigma},  \\
 \lambda_{\rm eff}&\equiv &\lambda_\phi-\frac{\lambda^2_{\sigma\phi}}{\lambda_\sigma}. 
 \eea
 Therefore, for $\phi^2\ll |m^2_\sigma/\lambda_{\sigma\phi}|$, a large effective non-minimal coupling $\xi_{\rm eff}$ appears for $\xi_2\gg 1$ \cite{gian} and the Higgs quartic coupling gets a tree-level shift due to the scalar threshold, curing the vacuum instability problem \cite{oleg-vsb,sthreshold}. 
In the limit of $\phi^2\gg |m^2_\sigma/\lambda_{\sigma\phi}|$, the Higgs field follows the sigma field satisfying $\sigma\approx \sqrt{-\lambda_{\sigma\phi} \phi^2/\lambda_\sigma}$. Then, for $\xi_2\gg \xi_\phi |\lambda_\sigma/\lambda_{\sigma\phi}|$, the effective frame function (\ref{efframe}) becomes
 \bea
 \Omega_{\rm eff}&\approx&1+\frac{\eta\sqrt{\phi^2}}{M_P}+\frac{\xi_{\phi,{\rm eff}} \phi^2}{M^2_P} 
 \eea
 with $\eta\approx \frac{\xi_1}{\sqrt{\xi_2}}\, \sqrt{\xi_{\phi,{\rm eff}}}$ and $\xi_{\phi,{\rm eff}}\approx -\frac{\lambda_{\sigma\phi}\xi_2}{\lambda_\sigma}>0$ for $\lambda_{\sigma\phi}<0$. 
As compared to the original Higgs inflation, the resulting frame function is augmented by a non-analytic form of the non-minimal coupling to gravity, $\sqrt{\phi^2} R$. As the sigma field theory with $\xi_1\sim \sqrt{\xi_2}$ is unitary up to the Planck scale, resulting in $\eta\sim \sqrt{\xi_{\phi,{\rm eff}}}$, the effective Higgs inflation keeps a similar prediction for inflation as in the sigma field theory.

\section{Inflaton dynamics with sigma field}

Now we discuss the general inflation vacua at large fields in our model. 
As far as the stable gravity condition (\ref{stablegravity}) is satisfied, the frame function $\Omega$ is always positive during inflation. For $\xi_1 \sigma+\xi_2\sigma^2+\xi_\phi\phi^2 \gg 1$ during inflation, we get $\Omega\approx\xi_1 \sigma+\xi_2\sigma^2+\xi_\phi\phi^2 $ and introduce a  new set of fields by 
\bea
e^{\frac{2}{\sqrt{6}} \chi}&=&\xi_1 \sigma+\xi_2\sigma^2+\xi_\phi\phi^2, \\
\tau&=& \frac{\phi}{\sigma}.
\eea
Then, since $e^{\sqrt{\frac{2}{3}}\chi} \gg \frac{\xi^2_1}{4(\xi_2+\xi_\phi\tau^2)}$ at large fields,  we get the approximate relation between $\sigma$ and redefined fields, $\chi$ and $\tau$,  as 
\bea
\sigma\approx \frac{e^{\frac{1}{\sqrt{6}}\chi}}{(\xi_2+\xi_\phi \tau^2)^{1/2}} \left(1-\frac{a}{2} \,e^{-\frac{1}{\sqrt{6}}\chi}+\frac{a^2}{8}\, e^{-\frac{2}{\sqrt{6}}\chi}\right)
\eea
with 
\be
a\equiv \frac{\xi_1}{(\xi_2+\xi_\phi \tau^2)^{1/2}}.
\ee
As a consequence, the scalar potential in Einstein frame becomes
\bea
V_E(\chi)&=& \frac{1}{4} (\lambda_\phi \tau^4+2\lambda_{\sigma\phi}\tau^2+\lambda_\sigma)\Big(1+e^{-\frac{2}{\sqrt{6}} \chi}\Big)^{-2} \sigma^4 \nonumber \\
&\approx&V_0 \left(1-2 a \,e^{-\frac{1}{\sqrt{6}}\chi} -(2+a^2) \, e^{-\frac{2}{\sqrt{6}}\chi} \right)  \label{inflapot}
\eea
with 
\bea
V_0\equiv \frac{\lambda_\phi \tau^4+2\lambda_{\sigma\phi}\tau^2+\lambda_\sigma}{4(\xi_2+\xi_\phi \tau^2)^2}.
\eea
Here, the condition of stable gravity (\ref{stablegravity}) requires that $|a|<2$. 

On the other hand, for $\xi_1\ll 2(\xi_2+\xi_\phi \tau^2)\sigma$, the kinetic terms for $\sigma$ and $\phi$ in eq.~(\ref{sigma-ein}) can be rewritten in terms of $\chi$ and $\tau$ \cite{oleg} as follows,
\bea
\frac{{\cal L}_{\rm kin}}{\sqrt{-g_E}}
&\approx& 
\frac{1}{2} \bigg[1+ \frac{1}{6} \frac{1+\tau^2}{\xi_2+\xi_\phi \tau^2} \bigg] (\partial_\mu\chi)^2 \nonumber \\
&&+\frac{1}{\sqrt{6}}  \frac{\tau(\xi_2-\xi_\phi)}{(\xi_2+\xi_\phi \tau^2)^2}\,(\partial_\mu\chi)(\partial^\mu\tau)\nonumber \\
&&+\frac{1}{2} \frac{\xi^2_\phi\tau^2+\xi^2_2}{(\xi_2+\xi_\phi \tau^2)^3} (\partial_\mu\tau)^2.  \label{kindiag}
\eea
Here, we note that for $\xi_2+\xi_\phi\gg 1$ or $\xi_2=\xi_\phi$ or $\langle\tau\rangle=0$, we can safely ignore the kinetic mixing term between $\chi$ and $\tau$. The results coincide with those in Ref.~\cite{oleg} for $\xi_1=0$.

We turn to the stabilization of $\tau$ from the scalar potential in eq.~(\ref{inflapot}).
Ignoring the third term in eq.~(\ref{inflapot}) in stabilizing $\tau$, for $\xi_2\gg \xi_\phi={\cal O}(1)$ and quartic couplings of order unity,  we get the conditions for the inflation vacua \cite{oleg} as
\begin{eqnarray}
&& (1)~\tau = \sqrt{ - \frac{\lambda_{\sigma\phi}}{\lambda_\phi}  }   \;: ~\lambda_\phi >0~, ~\lambda_{\sigma\phi} <0~, \nonumber\\
&& (2)~\tau=0  \;:~\lambda_\phi >0~,~
    \lambda_{\sigma\phi}  >0~, \nonumber\\
&& (3)~\tau=\infty  \;:~  \lambda_\phi <0~,~
\lambda_{\sigma\phi}  <0~,\nonumber\\
&& (4)~\tau=0,\infty  \;:~ \lambda_\phi  <0~,~
   \lambda_{\sigma\phi}  >0~.\label{taumin2}
\end{eqnarray}
In the first two cases, we need the Higgs quartic coupling to be positive during inflation: the former is the sigma-Higgs mixed inflation and the latter is the pure sigma inflation.  
In the third case, as the Higgs quartic coupling is required to be negative as $\lambda_\phi<0$, $V_0<0$, so it is not possible to get a dS vacuum for inflation. But, in the fourth case, even for $\lambda_\phi<0$, the inflation could be driven by the sigma field at the metastable vacuum with $\tau=0$ so it could lead to a viable cosmology with correct electroweak symmetry breaking at low energy. 
The resulting vacuum energy  for the viable inflation becomes
\bea
&& (1):~V_0=\frac{1}{4\xi^2_2} \Big(\lambda_\sigma-\frac{\lambda^2_{\sigma\phi}}{\lambda_\phi} \Big), \\
&& (2), \, (4):~V_0=\frac{\lambda_\sigma}{4\xi^2_2},
\label{taumin-b}
\eea
In all the above cases, then the sigma quartic self-coupling contributes dominantly to the inflaton vacuum energy. We note that the physical mass of the $\tau$ field is given as follows: (1)  $m^2_\tau\approx(-2\lambda_{\sigma\phi})/\xi_2$, or (2), (4):  $m^2_\tau\approx \lambda_{\sigma\phi}/\xi_2$, so $m^2_\tau\gg H^2\sim \lambda_\sigma/\xi^2_2$. Therefore, the dynamics of the $\tau$ field can be safely ignored during inflation.
The results are in agreement with the related analytic and numerical analyses for the effective single-field inflation in similar models as in Ref.~\cite{oleg,onefield}.

Consequently, from eqs.~(\ref{kindiag}) and (\ref{inflapot}),  we obtain the effective Lagrangian for the inflaton $\chi$, as follows,
\bea
\frac{{\cal L}_{\rm eff}}{\sqrt{-g_E}}= \frac{1}{2}R(g_E) - \frac{1}{2}(\partial_\mu\chi)^2 -V_E(\chi).  \label{infeff}
\eea
Therefore, for $a\gtrsim e^{-\frac{1}{\sqrt{6}}\chi}\sim 0.1$ during inflation, which makes the cutoff scale higher than $10^{-2}M_P$ even for a large $\xi_2\sim 10^4$, as shown in Fig.~\ref{cutoff}, the linear non-minimal coupling modifies the inflaton potential significantly, as compared with the case with quadratic non-minimal coupling only.

We remark that the linear non-minimal coupling $\xi_1$ in eq.~(\ref{fullag}) can be eliminated by redefining the $\sigma$ field with  ${\bar\sigma}= \sigma+\frac{\xi_1}{2\xi_2} M_P$, and then the frame function (\ref{frame}) and the scalar potential (\ref{fullpot}) in the Jordan frame are replaced by those with $\sigma={\bar\sigma}-\frac{\xi_1}{2\xi_2} M_P$.
Even in this case, small physical masses for the singlet scalar are kept.  The large VEV of the shifted field ${\bar\sigma}$ leads to a large kinetic mixing between the singlet scalar and the graviton in the Jordan frame or a large rescaling of the singlet scalar kinetic term in the Einstein frame \cite{gian,gian2}. 
For instance, for $\phi=0$ and $\xi_2 {\bar\sigma}^2\gg 1-\frac{\xi^2_1}{4\xi_2}$ during inflation, the canonical inflaton field becomes $\chi=\sqrt{\frac{3}{2}}\ln({\bar\sigma}^2/\xi_2)$ for the ${\bar\sigma}$ field. Then, from the $\lambda_\sigma$ term in this field basis, we obtain the same effective Lagrangian for the inflaton with $\tau=0$ as eq.~(\ref{infeff}).

\begin{figure}
  \begin{center}
    \includegraphics[height=0.40\textwidth]{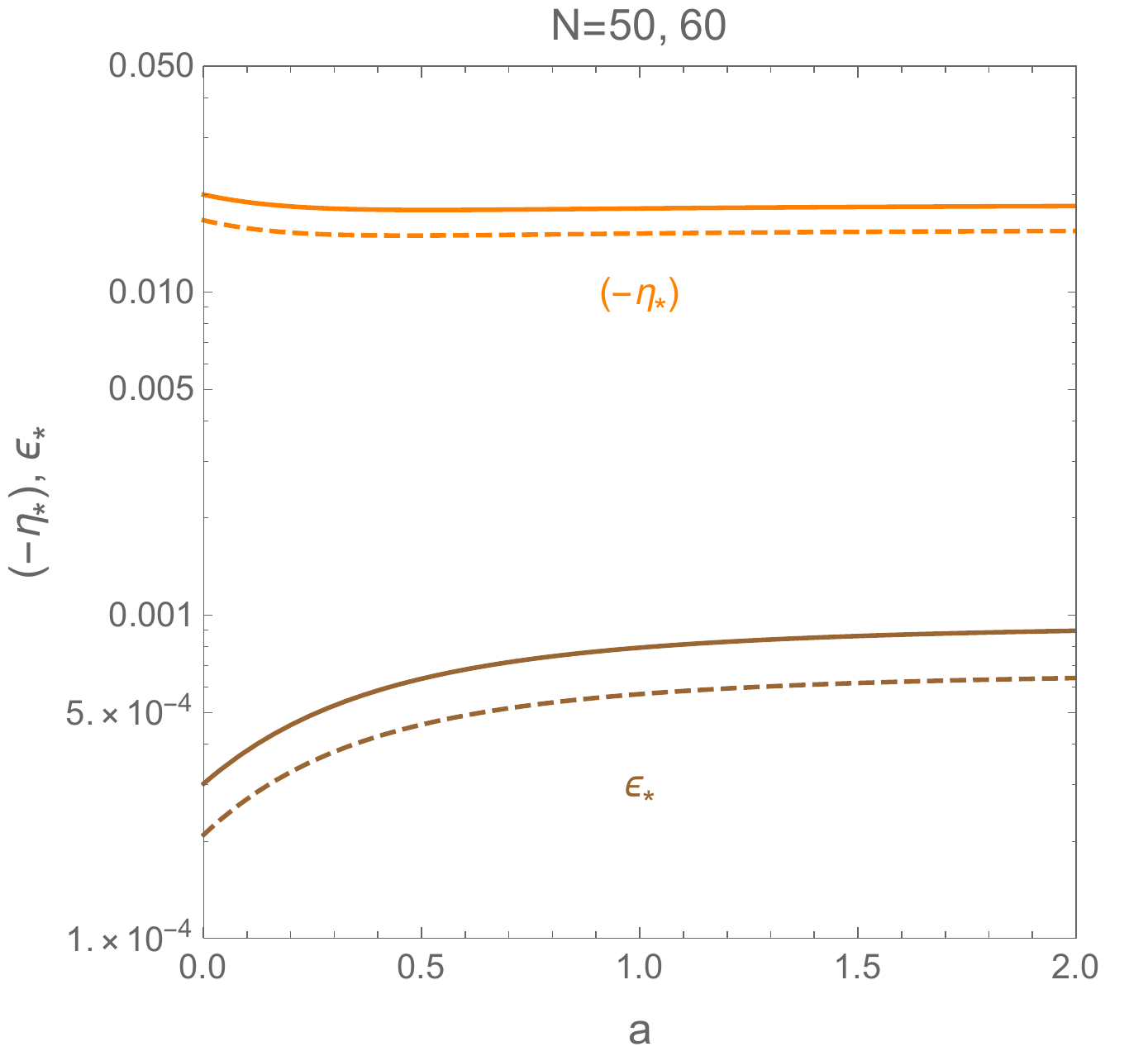}
  \end{center}
  \caption{Slow-roll parameters as a function of $a\equiv \xi_1/(\xi_2+\xi_\phi \tau^2)^{1/2}$. We have chosen $N=50, 60$ in solid and dashed lines, respectively. }
  \label{slowroll}
\end{figure}

From the effective inflaton Lagrangian in eq.~(\ref{infeff}),  the slow-roll parameters during inflation are given by
\bea
\epsilon&=& \frac{1}{3}(2+a^2)^2 \, e^{-\frac{2}{\sqrt{6}}\chi} \Big(e^{-\frac{1}{\sqrt{6}}\chi} +\frac{a}{2+a^2} \Big)^2,   \label{epsilon}\\
\eta&=&   -\frac{2}{3}(2+a^2) \, e^{-\frac{1}{\sqrt{6}}\chi} \Big(e^{-\frac{1}{\sqrt{6}}\chi} +\frac{a}{2(2+a^2)}\Big). \label{eta}
\eea
As a result, the spectral index is given by
$n_s =1-6\epsilon_*+ 2\eta_*$
where $*$ denotes the evaluation of the slow-roll parameters, (\ref{epsilon}) and (\ref{eta}), at horizon exit.
The tensor-to-scalar ratio is also given by $r=16\epsilon_*$ with eq.~(\ref{epsilon}) at horizon exit.
The measured spectral index and the bound on the
tensor-to-scalar ratio are given by $n_s=0.9652\pm 0.0047$ and $r < 0.10$ at $95\%$ C.L., respectively, from Planck TT, TE, EE + low P \cite{planck}.

\begin{figure}
  \begin{center}
    \includegraphics[height=0.40\textwidth]{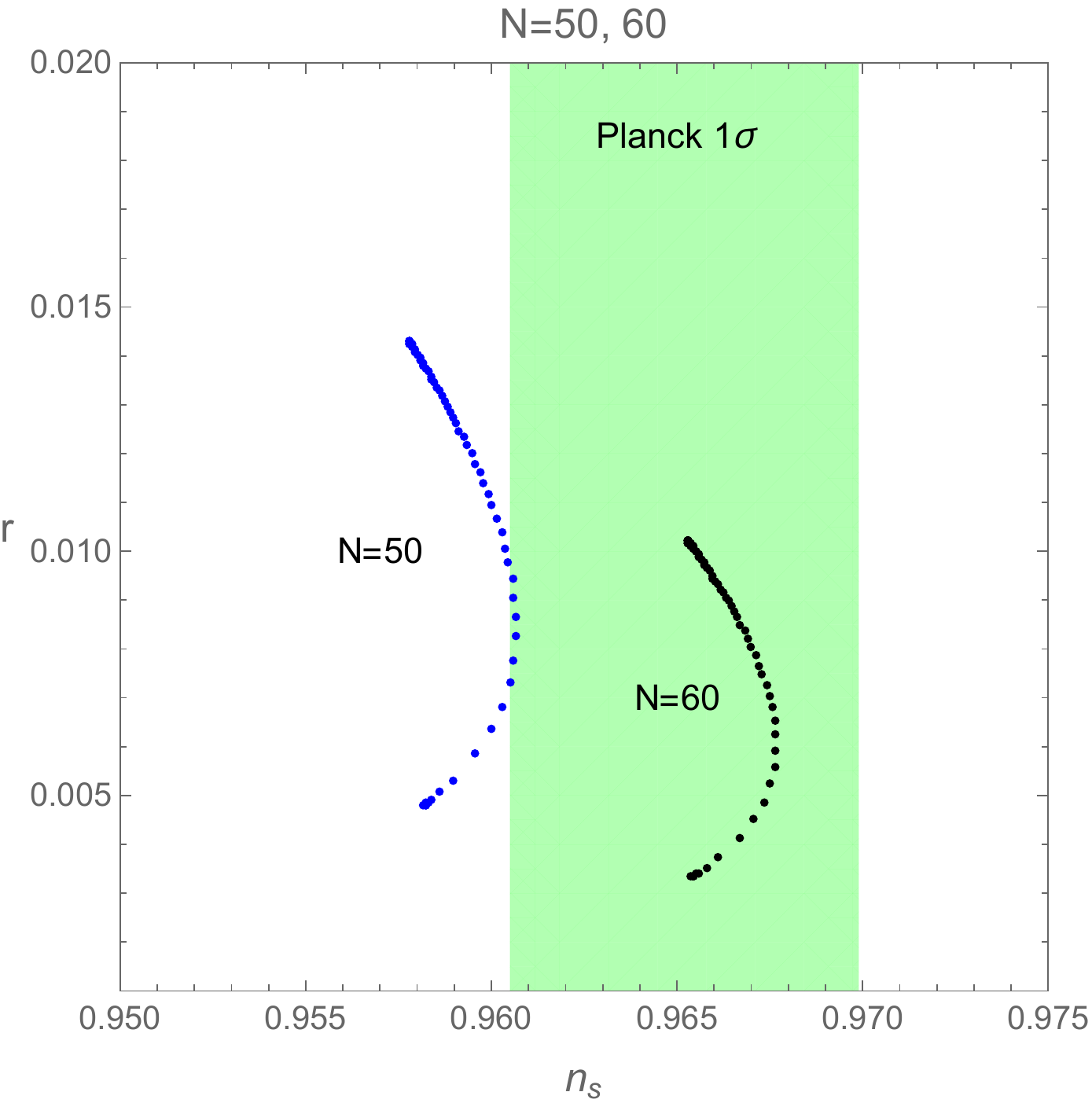}
  \end{center}
  \caption{Spectral index $n_s$ vs tensor-to-scalar ratio $r$ for $a=0-2$. We have chosen $N=50, 60$ in blue and black lines, respectively. Planck $1\sigma$ band is shown in green.}
  \label{ns-r}
\end{figure}

Moreover, the number of efoldings required to solve the horizon problem can be calculated as follows,
\bea
N&=& \int^{\chi_i}_{\chi_f} \frac{d\chi}{\sqrt{2\epsilon}} \nonumber \\
&\approx & \frac{3(2+a^2)}{a^2} \bigg[ \frac{a}{2+a^2} \, e^{\frac{1}{\sqrt{6}}\chi_*} - \ln \Big(1+ \frac{a}{2+a^2}\, e^{\frac{1}{\sqrt{6}}\chi_*} \Big)\bigg]
\eea
where 
$\chi_{i,f}$ are the inflaton values at the beginning and end of inflation and we can take $\chi_i=\chi_*$. In Fig.~\ref{slowroll}, we show the slow-roll parameters evaluated at the horizon exit as a function of $a$ for $N=50, 60$.  Thus, we find that $\eta_*$ is insensitive to $a$ but $\epsilon_*$ changes to a large value as $a$ approaches unity.

For $a\ll e^{-\frac{1}{\sqrt{6}}\chi_*} $, i.e. $\xi_1\ll 1$, the results with quadratic non-minimal couplings only are recovered, namely, $N\approx\frac{3}{4}\,  e^{\frac{2}{\sqrt{6}}\chi_*}$, $\epsilon_*\approx \frac{4}{3}\,  e^{-\frac{4}{\sqrt{6}}\chi} $ and $\eta_*\approx -\frac{4}{3}\,  e^{-\frac{2}{\sqrt{6}}\chi}$. Then, we get $\epsilon_*\approx \frac{3}{4N^2}$ and $\eta_*\approx -\frac{1}{N}$, so the spectral index and the tensor-to-scalar ratio become $n_s\approx 1-\frac{2}{N}$ and $r\approx \frac{12}{N^2}$, respectively \cite{Higgsinf}.

However, for a sizable $a$ or $\xi_1$, the inflationary observables are modified by the linear non-minimal coupling for $\sigma$.  
In Fig.~\ref{ns-r}, we depict  the correlation between the spectral index and the tensor-to-scalar ratio the spectral index for a varying $a\equiv  \xi_1/(\xi_2+\xi_\phi \tau^2)^{1/2}$ for $N=50, 60$ in blue and black lines, respectively. As $\epsilon_*$ increases for a sizable $a$, the tensor-to-scalar ratio  the tensor-to-scalar ratio varies significantly from the one for the original Higgs inflation \cite{Higgsinf} up to  $r=0.015$ for $N=50$ and $r=0.010$ for $N=60$. Thus, the primordial gravity waves are at the detectable level in the future CMB experiments such as LiteBIRD~\cite{future}.

\section{Conclusions}

We have presented the inflation model  with a light singlet field containing both linear and quadratic non-minimal couplings to gravity.  We showed that the inflaton potential is determined by the quartic couplings and the non-minimal couplings only. We found that the linear non-minimal coupling for the singlet field makes the model unitary up to Planck scale and allows for a sizable deviation in tensor-to-scalar ratio from Higgs inflation. The light singlet inflaton can be probed by low-energy phenomena such as the direct production at the Large Hadron Collider through Higgs interactions.

\textit{Acknowledgments}--- 
I would like to thank Gian F. Giudice for  valuable discussion. 
The work is supported in part by Basic Science Research Program through the National Research Foundation of Korea (NRF) funded by the Ministry of Education, Science and Technology (NRF-2016R1A2B4008759). 

\end{document}